\documentclass[12pt]{article}
\usepackage[dvips]{graphicx}
\def\thefootnote{\fnsymbol{footnote}}
\def\be{\begin{equation}}
\def\ee{\end{equation}}
\def\ba{\begin{eqnarray}}
\def\ea{\end{eqnarray}}

\begin{document}
\begin{titlepage}
\thispagestyle{empty}
\vskip0.5cm
\vskip0.8cm

\begin{center}
{\Large {\bf Minimal Variance Hedging of Options}}
\end{center}
\begin{center}
{\Large {\bf with Student-t Underlying}}
\end{center}
\vskip0.8cm
\begin{center}
{\large Klaus Pinn}\\
\vskip5mm
{Institut f\"ur Theoretische Physik I } \\
{Universit\"at M\"unster }\\ {Wilhelm--Klemm--Str.~9 }\\
{D--48149 M\"unster, Germany \\[5mm]
 e--mail: pinn@uni--muenster.de
 }
\end{center}
\vskip1.5cm
\begin{abstract}
\par\noindent
I explicitly work out closed form solutions for 
the optimal hedging strategies (in the sense of
Bouchaud and Sornette)  
in the case of European call
options, where the underlying is modeled by 
(unbiased) iid additive returns with
Student-t distributions. The results may serve as illustrative
examples for option pricing in the presence of fat tails. 

\end{abstract}
\end{titlepage}

\setcounter{footnote}{0}
\def\thefootnote{\arabic{footnote}}

\section{Introduction}

Black and Scholes option pricing theory relies on the existence of a
dynamic trading strategy in the underlying asset that exactly
replicates the option holder's claim [1-4].
There is no risk for the writer, and the fair option price can
be interpreted as the cost of the perfect 
hedging strategy.

Real world markets are, however, not complete, i.e., there are no
riskless hedging strategies that would provide unique option prices.
The incompleteness stems from a number of reasons. E.g., stock market
returns are not log-normally distributed (as assumed in the Black
Scholes analysis). It is widely agreed that marginal return
distributions have fat tails, i.e., large events occur with
significantly higher frequency than to be expected from the log-normal
assumption.

An interesting approach to option theory for incomplete markets has
been worked out by M\"uller \cite{Mueller}, F\"ollmer and Sondermann
\cite{Foellmer}, Sch\"al \cite{Schael} and Schweizer \cite{Schweizer},
and also by Bouchaud and coworkers \cite{Bouchaud1,Bouchaud2}, see
also [11-15].  At the heart of these approaches lies a minimization of
the risk inherent in writing the option.  Different risk measures have
been suggested, including the global and local variance of the cost
process [5-7] and the variance of the global operator
wealth~\cite{Bouchaud1}.

In this article, I will explicitly work out closed form solutions for
the optimal hedging strategies (in the sense of Bouchaud and Sornette)
in the case of European call options, where the underlying is modeled
by (unbiased) iid additive returns with Student-t distributions. The
results may serve as illustrative examples for option pricing in the
presence of fat tails.  Furthermore, since Student distributions have
been considered as acceptable models of market returns [16-20], 
the results presented here could also be useful for practical purposes.

This article is organized as follows: In section 2, I setup the model
and notation. A number of useful formulae is derived that are of
significant help in the explicit calculations of later sections.  In
section 3 we reproduce a few (well known) results for Gaussian
returns. In section 4 I derive the closed form
solutions\footnote{Closed form here means that the result is presented
as a finite sum of terms built from standard functions.} for the
Student-t case.  Illustrative example results are given in section 5.
Conclusions follow.  Some technical parts are deferred to three
appendices.

\section{Model Setup and Basic Formulae} 

Consider a European call option, with maturity at $T=N\tau$, 
and strike price $x_s$. Trading and hedging is only 
possible at discrete times $t= k \, \tau$, $k= 0,\dots,N$. 
The underlying price process is given by $x_k \equiv x(k \, \tau)$.  
At expiry the option holder's claim is given by 
$$
f(x_N) = {\rm max}(x_N - x_s,0) \, . 
$$
For the sake of 
simplicity we assume that the (risk free) interest rate is 
zero. 

After writing an option, the bank will perform the following 
(hedging) actions. $k= 0,\dots,N-1$: 
Trade the underlying such that the bank's 
portfolio contains $\phi_k$ pieces of stock. 
$k= N$: 
Sell the portfolio and satisfy the option holder's claim. 

Following Bouchaud and Sornette \cite{Bouchaud1}, we write down the 
global cost balance for the bank 
(per option issued):\footnote{We have not included the 
option premium in the balance.}
\be
\label{balance}
C = {\rm max}(x_N - x_s,0) 
- \sum_{k=0}^{N-1} \phi_k(x_k) \, (x_{k+1} - x_k) \, . 
\ee
If it were predictable (deterministic), $C$ would be the correct price
of the option (ignoring transaction action costs etc.).  
However, as a function of the underlying, $C$ is a stochastic variable.
The fluctuations of $C$ will in general 
depend on the choice of the hedging strategy $\phi$. 
In the Bouchaud Sornette approach, on chooses $\phi$ in order 
to make $C$ as determininistic as possible, e.g., by  
demanding that the variance of $C$ is minimal.
The optimal hedging strategy $\phi^*$ is then determined by 
\be 
R^2(\phi^*)= \langle C^2 \rangle - 
\langle C \rangle^2  = {\rm minimal} \, , 
\ee 
where $\langle . \rangle$ denotes averaging over the price process. 
For sufficiently small residual risk, e.g., 
for $R/\langle C \rangle < \!\!< 1$, it is reasonable 
for the bank to define the option price as 
$$
{\rm price} = \langle C \rangle + 
\mbox{premium for the residual risk} \, . 
$$
The extra premium for the residual risk  
will of course depend on the bank's risk preferences. 

Let us now assume the following (simplified) model for the 
returns of the underlying asset: 
$$
x_{k+1} - x_k = r_{k+1} \, , \quad k=0,\dots,N-1 \, , 
$$ 
where the $r$'s are iid with probability density function
(pdf) $p(r)$.
We require that $p$  
should have a finite variance, defined 
by\footnote{Integrals without explicit limits should always 
be extended over the whole real axis.}
$$
\sigma^2 = \int dr \, p(r) \, r^2 \, . 
$$
However, we are not assuming that 
$p$ is Gaussian or log-normal. 
The stock price $x_k$ can be written as 
$$
x_k = x_0 + \sum_{i=1}^k r_k \equiv x_0 + R_k \, . 
$$
The probability measure is defined by 
$$
\langle F \rangle = 
\int dr_1\, p(r_1) \dots \int dr_N \, p(r_N) \, 
F(r_1,\dots,r_N) \, . 
$$
We will assume that $p(r)$ is even, i.e., $p(r)= p(-r)$.
Then the expectation value of $C$ is unaffected 
by the hedging strategy $\phi$: 
$\langle C \rangle = \langle f \rangle$. 
In the present context it is given by 
$$
\langle C \rangle = 
\int_{-\infty}^{\infty} 
dt \, P_N(t) \, {\rm max}(x_0 + t - x_s,0) 
= \int_{-\zeta}^{\infty} 
dt \, P_N(t) (\zeta + t) 
$$
We have defined  
$$
\zeta = x_0 - xs \, . 
$$
At $k=0$ the option is in-the-money, at-the-money, or 
out-of-the-money, for $\zeta>0$, $\zeta =0$, or $\zeta<0$, respectively. 

$P_l$ is the pdf of $R_l$, i.e., the 
$l$-fold convolution of $p$. It can be conveniently accessed 
via Fourier transformation methods. 
Introducing the characteristic function $\tilde p(q)$
through 
$$
\tilde p(q) = \int dr \, e^{iqr} \, p(r) \, , 
$$
we have 
$$
\tilde P_l(q) = \tilde p(q)^l \, . 
$$

It is not difficult to show that the hedging strategy minimizing 
the variance of $C$ can be expressed as 
$$ 
\phi_k^*(x_k) = 
\frac{1}{\sigma^2} \int dr \, p(r) \, r 
\int dt \, P_{N-k-1}(t) \, f(x_k + r + t) \, . 
$$
It is demonstrated in Appendix A that 
\be
\label{psiform} 
\psi_k^* = - \frac{1}{(N-k) \, \sigma^2} 
\int \frac{dq}{2\pi} \, 
\frac{\sin(q \, \xi)}{q} \, 
\, \frac{1}{q} \, \frac{d}{dq} \, \tilde p(q)^{N-k} \, , 
\ee 
where $\phi_k^* = \frac12 + \psi_k^*$, and 
$\xi \equiv x_k - x_s$.

The residual risk can be expressed as 
\be 
\label{Rrisk}
R^2(\phi^*) = 
\langle f^2 \rangle - 
\langle f \rangle^2 
- \sigma^2 \, \sum_{k=0}^{N-1} \int dx \, 
P_k(x-x_0) \, \phi_k^*(x)^2 \, .  
\ee 
Here, 
$ \langle f^2 \rangle - \langle f \rangle^2 $ is the unhedged risk. 
$\langle f \rangle$ and 
$\langle f^2 \rangle$ can be computed with the help 
of the following relations, derived in Appendix B: 
\be 
\label{LL1}
\begin{array}{ll}
\langle f^{\phantom{1}} \rangle &=
\zeta \, \left(  
\frac{1}{2} + L_1(\zeta) \right) + L_2(\zeta) \, , 
\\[4mm] 
\langle f^2 \rangle &= 
\zeta^2 \, \left(  
\frac{1}{2} + L_1(\zeta) \right) 
+ \zeta \, L_2(\zeta) + 
\left( \frac12 N\, \sigma^2 + L_3(\zeta) \right) \, , 
\end{array}
\ee 
where 
\be 
\label{LL2}
\begin{array}{ll}
L_1(\zeta) &= \int \frac{dq}{2\pi} \, \frac{\sin(\zeta q)}{q}
\, \tilde p(q)^N \, , 
\\[4mm] 
L_2(\zeta) &= - \int \frac{dq}{2\pi} \, \cos( \zeta q) 
\, \frac{1}{q} \, \frac{d}{dq} \, \tilde p(q)^N \, , 
\\[4mm] 
L_3(\zeta) &= - \int \frac{dq}{2\pi} \, \frac{\sin(\zeta q)}{q}
\, \frac{1}{q} \, \frac{d}{dq} \, \tilde p(q)^N \, .
\end{array}
\ee 
The second term in eq.~(\ref{Rrisk}) can be rewritten as 
$$
\sigma^2 \sum_{k=0}^{N-1} 
\int d\xi \, P_k(\xi-\zeta) \, 
\left( \frac12 + \psi_k^*(\xi) \right)^2 \, . 
$$
In the examples to be discussed below we shall evaluate this term
by performing the $\xi$-integration numerically, starting 
from closed form expressions for $P_k$ and $\psi^*_k$.

\section{The Gaussian Case}

As a first application of the formulae derived in the previous section, 
and for later comparison with the Student-t case, 
let us consider the case of Gaussian returns, i.e.,
$ p(r) = 1/\sqrt{2\pi \sigma^2} \, 
e^{-r^2/(2\sigma^2)}$. 
The characteristic function is $\tilde p(q)=e^{-\frac12 \sigma^2 q^2}$.
The integrals for $\psi_k^*$ and $L_i$ are elementary. 
With $\sigma_l^2= l \, \sigma^2$ we find 
\be
\label{gaussh}
\psi_{k,G}^* = \frac{1}{2} \, 
{\rm erf}\left( \xi / \sqrt{2\sigma_{N-k}^2}  \right) \, . 
\ee
For $\phi_{k,G}^* = \frac12 + \psi_{k,G}^*$ one thus obtains 
$ \phi_{k,G}^* = \hat N(\xi/\sigma_{N-k})$,
where $\hat N$ denotes the cumulative distribution function 
of the normal distribution. For the $L_i$ we get 
$$
\begin{array}{ll}
L_{1,G} &= \frac12 \, 
{\rm erf}\left( \zeta / \sqrt{2\sigma_N^2}  \right) \, , \\[4mm]
L_{2,G} &= \sigma_N^2 \, 
\frac{e^{-\zeta^2/(2 \sigma_N^2)}} {\sqrt{2\pi \sigma_N^2}} \, , \\[4mm]
L_{3,G} &= \frac12 \, \sigma_N^2 \,  
{\rm erf}\left( \zeta / \sqrt{2\sigma_N^2}  \right) \, . \\[4mm]
\end{array}
$$
For a complete discussion of the Gaussian case see 
\cite{Bouchaud1,Bouchaud2}.


\section{Options with Student-t Underlying}

We shall now consider the case that the return distribution is of the 
Student-t type, 
$$
p^{(\mu)}(r) = 
\frac{\Gamma\left(\frac12(1+\mu)\right)}
{\sqrt{\pi} \, \Gamma\left(\frac12 \mu\right)}
\, 
\frac{a^\mu}
{(a^2+r^2)^{(1+\mu)/2}} \, . 
$$
$a > 0$ is a scale parameter that we shall later use 
to fix the standard deviation. The latter exists for 
$\mu > 2$.
For $\mu$ an odd integer, 
one can derive relatively simple 
expressions for the 
characteristic functions $\tilde p^{(\mu)}$, 
thus making feasible closed form solutions 
for the hedge functions $\phi^*_k$. 
Let us therefore concentrate on the case $\mu = 2\, m + 1$, with 
$m= 1,2,\dots$. By an elementary integration one can show that 
\be
\label{ssig}
\sigma^2 = \frac{a^2}{2m-1} \, . 
\ee
A straightforward but somewhat lengthy calculation (employing 
the theorem of residues) 
yields 
$$
\tilde p^{(2m+1)}(q) = 
e^{-a|q|} \, T^{[m]}(a|q|) \, ,   
$$
where 
$$
T^{[m]}(q) = \sum_{k=0}^{m} 
\left(  
\begin{array}{c}
m \\
k 
\end{array}
\right) 
\left/
\left(  
\begin{array}{c}
2 m \\
k 
\end{array}
\right) 
\right.
\, 
\frac{(2\,q)^k}{k!}
$$
is a polynomial of degree $m$.  
Here are a few examples: 
$$
\begin{array}{ll}
T^{[0]}(q) &= 1 \, , \\[3mm]
T^{[1]}(q) &= 1 + q \, , \\[3mm]
T^{[2]}(q) &= 1 + q + \frac13 q^2 \, .
\end{array}
$$
In Appendix C eq.~(\ref{psiform}) is used to derive a 
closed form expression for $\psi_k^{[m]*}$. 
The result is 
$$
\psi_k^{[m]*} = \frac1\pi \arctan \frac{\xi'}{N-k}
+ \sum_{l=1}^{m(N-k)-1} \frac{A_l^{[m,N-k]} \, (l-1)!}{\pi} \, 
\frac{ \sin\left(l \, \arctan\left(\frac{\xi'}{N-k}\right)\right)}
{\left(\xi'^2 + (N-k)^2\right)^{l/2}} \, . 
$$
Here we have defined 
$$
\xi' = \frac{\xi}{a} = \frac{x_k-x_s}{a} \, . 
$$
The coefficients $A^{[m,N-k]}_j$ are defined by 
$$ 
T^{[m]}(q)^{N-k-1} \,  
T^{[m-1]}(q) \equiv \sum_{j=0}^{m(N-k)-1} A^{[m,N-k]}_j \, q^j \, .
$$
Calculations very similar to those presented in Appendix C 
help us to find explicit expressions for the $L_i$, defined 
in eq.~(\ref{LL2}).

For $L_1$ we get 
$$
L_1^{[m]} = \int_0^\infty \frac{dq}{\pi} \, 
\frac{\sin(\zeta'q)}{q} \, 
e^{-Nq} \, T^{[m]}(q)^N \, , 
$$
with 
$\zeta' = \zeta/a$.
Expanding 
$$
T^{[m]}(q)^N = \sum_{l=0}^{mN} B_l^{[m,N]} \, q^l \, , 
$$
we find 
$$
L_1^{[m]} = \frac1\pi \arctan \frac{\zeta'}{N}
+ \sum_{l=1}^{mN} \frac{B_l^{[m,N]} \, (l-1)!}{\pi} \, 
\frac{ \sin\left(l \, \arctan\left(\frac{\zeta'}{N}\right)\right)}
{\left(\zeta'^2 + N^2\right)^{l/2}} \, .
$$
Similarly, for $L_2$ one obtains 
$$
L_2 = 
\frac{N\sigma^2}{a} \int_0^\infty  
\frac{dq}{\pi} \, \cos(\zeta' q) \, 
e^{-Nq} \, 
T^{[m]}(q)^{N-1} \,  T^{[m-1]}(q) \, . 
$$
Using that 
$$
\int_0^\infty 
dx \, 
x^{\alpha-1} \, e^{-\beta \,x} \, 
\cos( \gamma x) = \frac{\Gamma(\alpha)}{(\beta^2+\gamma^2)^{\alpha/2}}
\, \cos\left( \alpha \, \arctan \frac\gamma\beta \right) \, , 
$$
one arrives at 
$$
L_2^{[m]} = \frac{N \sigma^2}{a}
\sum_{l=0}^{mN-1} \frac{A_l^{[m,N]} \, l!}{\pi} \, 
\frac{ \cos\left((l+1) \, \arctan\left(\frac{\zeta'}{N}\right)\right)}
{\left(\zeta'^2 + N^2\right)^{(l+1)/2}} \, .
$$
The calculation of $L_3$ goes along the same lines. We find 
$$
L_3^{[m]} = N\sigma^2 
\left( \frac1\pi \arctan \frac{\zeta'}{N}
+ \sum_{l=1}^{mN-1} \frac{A_l^{[m,N]} \, (l-1)!}{\pi} \, 
\frac{ \sin\left(l \, \arctan\left(\frac{\zeta'}{N}\right)\right)}
{\left(\zeta'^2 + N^2\right)^{l/2}} 
\right) \, .
$$
For the calculation of the residual risk we
shall also need an expression for $P_k$, the $k$-fold convolution 
of $p$. The result is 
$$
a \, P_k(at) = 
\sum_{l=0}^{mk} \frac{B_l^{[m,k]} \, l!}{\pi} \, 
\frac{ \cos\left((l+1) \, \arctan\left(\frac{t}{k}\right)\right)}
{\left(t^2 + k^2\right)^{(l+1)/2}} \, .
$$


\section{Example Results}

\begin{figure}
\begin{center}
\includegraphics[width=10cm]{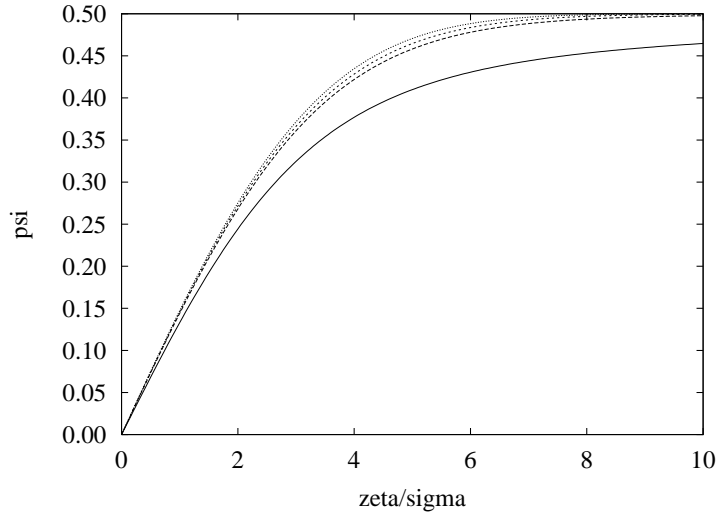}
\parbox[t]{.85\textwidth}
 {
 \caption[fig1]
 {\label{fig1}
\small
The $N=7$, $k=0$ 
hedging function $\psi_0^*$ for Student-t underlying 
with $m=1$, 2, and 3 (bottom to top). The uppermost curve 
is the hedge function for the Gaussian case. 
 }
 }
\end{center}
\end{figure}

In this section we will present some example results for 
the Student-t distribution, in comparison with the 
Gaussian case. 
Let us start with a look at the hedge functions for 
different values of $m$. We choose $N=7$ and $k=0$. 
If you think of days as the hedging and trading intervals,
this corresponds to one week before expiry.
Figure~\ref{fig1} shows the results for 
$m=1$, 2, and 3, together with the Gaussian results. 
$\psi^*_0$ is plotted as function of $\zeta/\sigma$, where 
$\sigma$ denotes the standard deviation (per day).

\begin{figure}
\begin{center}
\includegraphics[width=10cm]{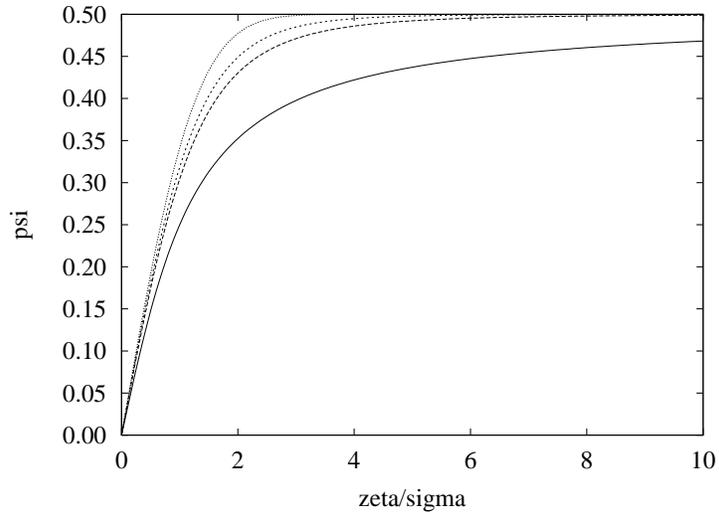}
\parbox[t]{.85\textwidth}
 {
 \caption[fig2]
 {\label{fig2}
\small
The $N=1$, $k=0$ 
hedging function $\psi_0^*$ for Student-t underlying 
with $m=1$, 2, and 3 (bottom to top). The uppermost curve 
is the hedge function for the Gaussian case. 
 }
 }
\end{center}
\end{figure}
With increasing $m$, the curves converge quickly towards the 
$\psi$-function of the Gaussian returns. Closer to expiry 
the difference to the Gaussian behaviour is more significant, 
as demonstrated in Figure~\ref{fig2}, which shows $\psi^*_0$
for the same set of $m$-values, however with $N-k=1$, i.e.,
one day before maturity. 

\begin{figure}
\begin{center}
\includegraphics[width=10cm]{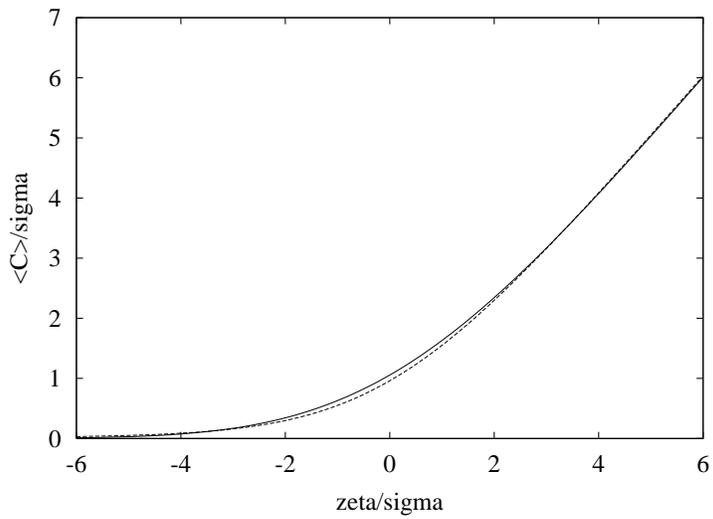}
\parbox[t]{.85\textwidth}
 {
 \caption[fig3]
 {\label{fig3}
\small
$\langle C \rangle$ for 
$N=7$, Student-t, with $m=3$, in comparison with the 
Gaussian.  The Student-t result is below the Gaussian for small $\zeta$. 
 }
 }
\end{center}
\end{figure}

Let us next consider the behaviour of $\langle C \rangle$. 
Figure~\ref{fig3} shows this quantity for $N=7$, for 
$m=3$, and for the Gaussian,
as a function of $\zeta/\sigma$. Around $\zeta=0$, 
the Student result is a little bit below the Gaussian, farther 
out it is a very little bit above it (``volatility smile''). 

\begin{figure}
\begin{center}
\includegraphics[width=10cm]{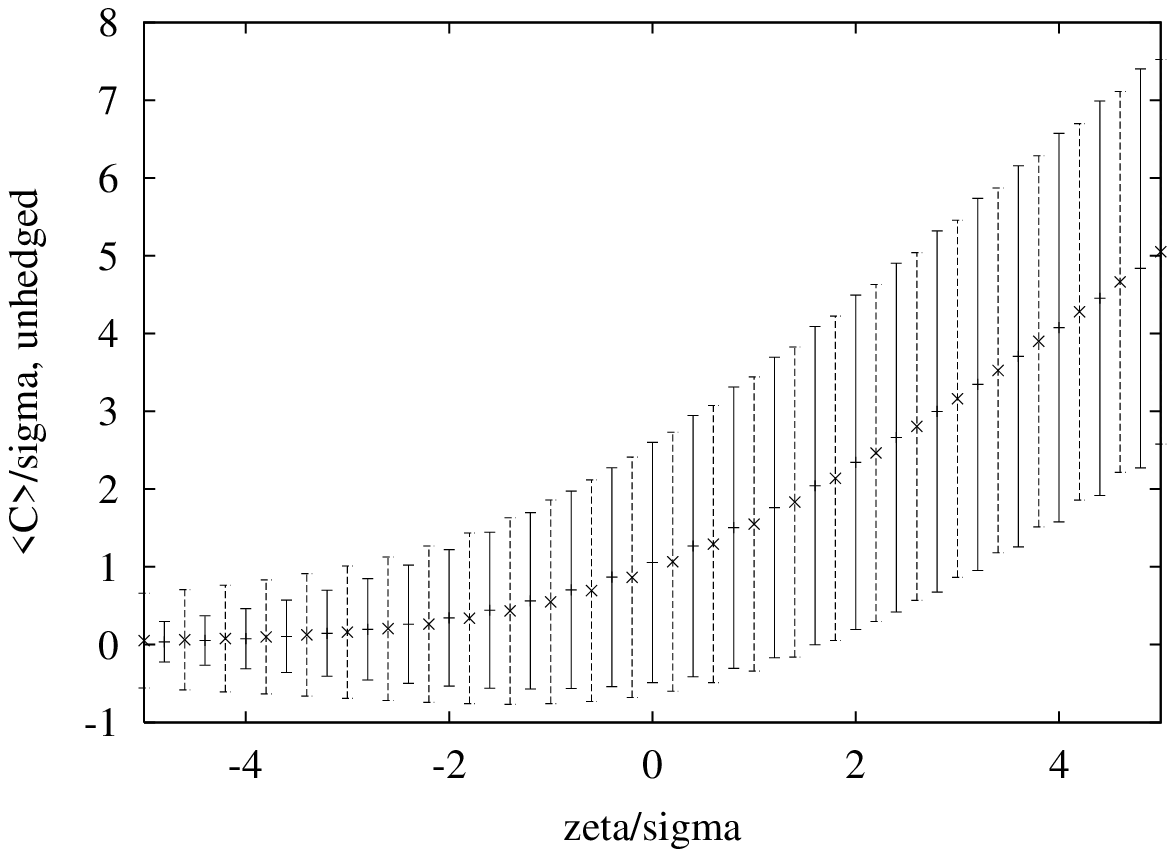}
\includegraphics[width=10cm]{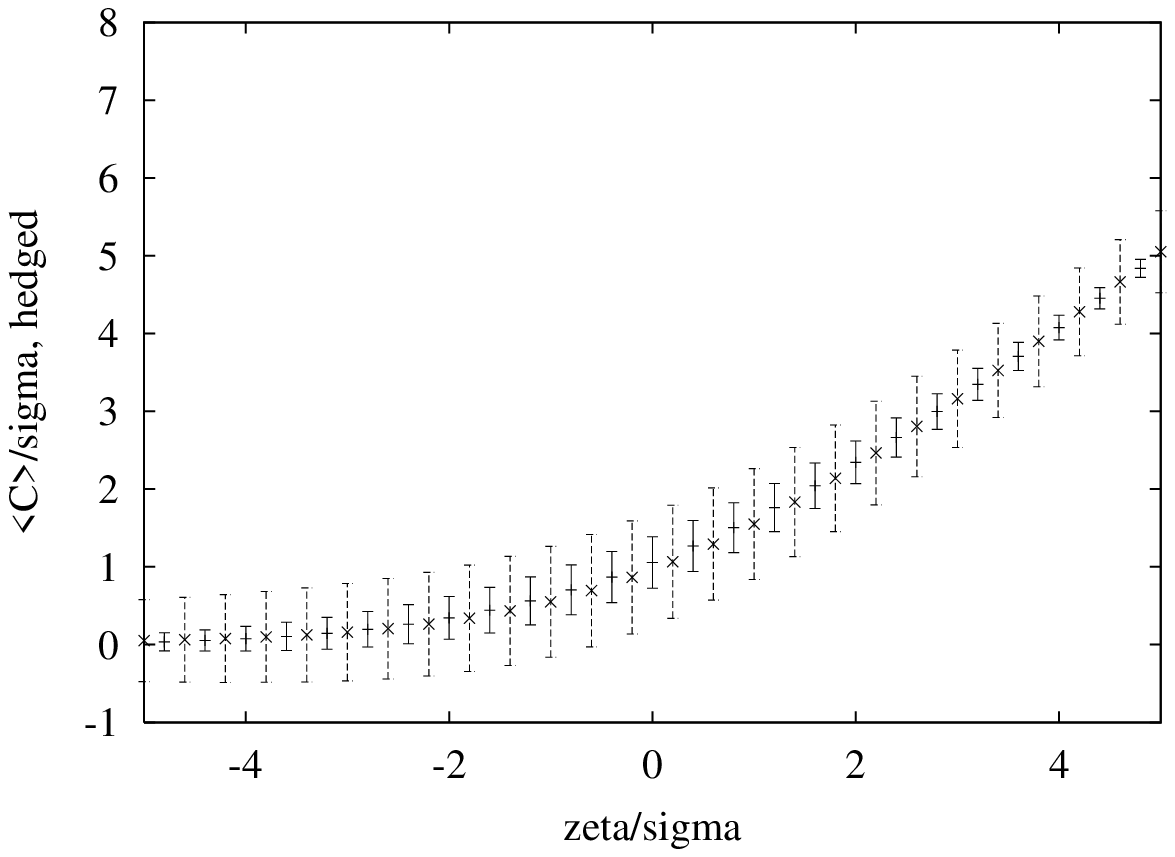}
\parbox[t]{.85\textwidth}
 {
 \caption[fig4]
 {\label{fig4}
\small
For $N=7$: $\langle C \rangle$, together with 
$\langle C^2\rangle - \langle C \rangle^2$, 
plotted with the help of error bars.  
Top: unhedged, bottom: optimally hedged. Bars: 
Gaussian, crosses: Student-t with $m=1$.
 }
 }
\end{center}
\end{figure}

Figure~\ref{fig4} shows again results for $N=7$, $\langle C \rangle$ as 
a function of $\zeta/\sigma$, together with the risk 
$\langle C^2\rangle - \langle C \rangle^2$, 
plotted with the help of error bars.  
The upper plot gives these quantities for the unhedged case, 
i.e., for $\phi=0$. For the lower plot the optimal hedging 
was performed. 
In both cases, the data with the smaller risk belong to 
the Gaussian, the other ones to the Student-t with $m=1$. 
Note that there remains quite some risk even after the 
optimal hedging. It is only for the Gaussian that this 
residual risk 
can be removed in the limit of continous hedging. 

\begin{figure}
\begin{center}
\includegraphics[width=10cm]{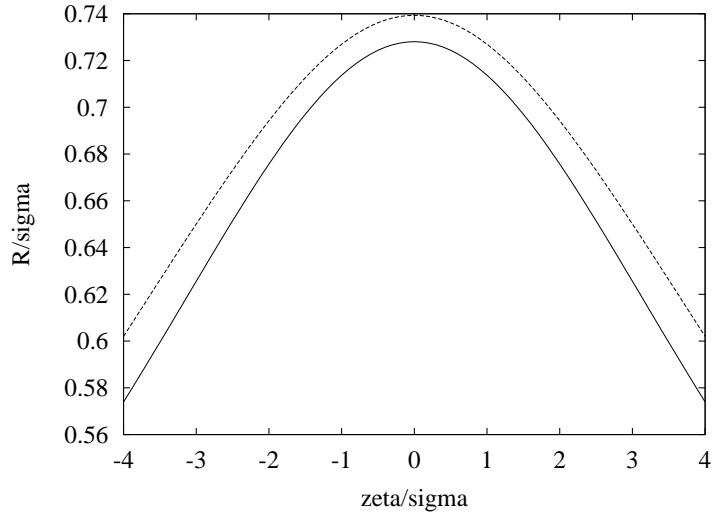}
\parbox[t]{.85\textwidth}
 {
 \caption[fig5]
 {\label{fig5}
\small
Upper curve: residual 
risk for $N=7$, Student-t with $m=1$, hedged 
with Gaussian hedge function.
The lower curve is the 
residual risk in the case of optimal hedging. 
 }
 }
\end{center}
\end{figure}

It is interesting to visualize the ``hedging error'' occuring 
when one hedges the Student-t underlying with the 
Gaussian hedge functions given by eq.~(\ref{gaussh}).
The upper curve in Figure~\ref{fig5} denotes the residual 
risk for $N=7$, Student-t returns with $m=1$, hedged 
with the Gaussian hedge function (corresponding to the 
same standard deviation). The lower curve is the 
residual risk in the case of optimal hedging. 
The difference is not dramatic, but appreciable. 

\begin{figure}
\begin{center}
\includegraphics[width=10cm]{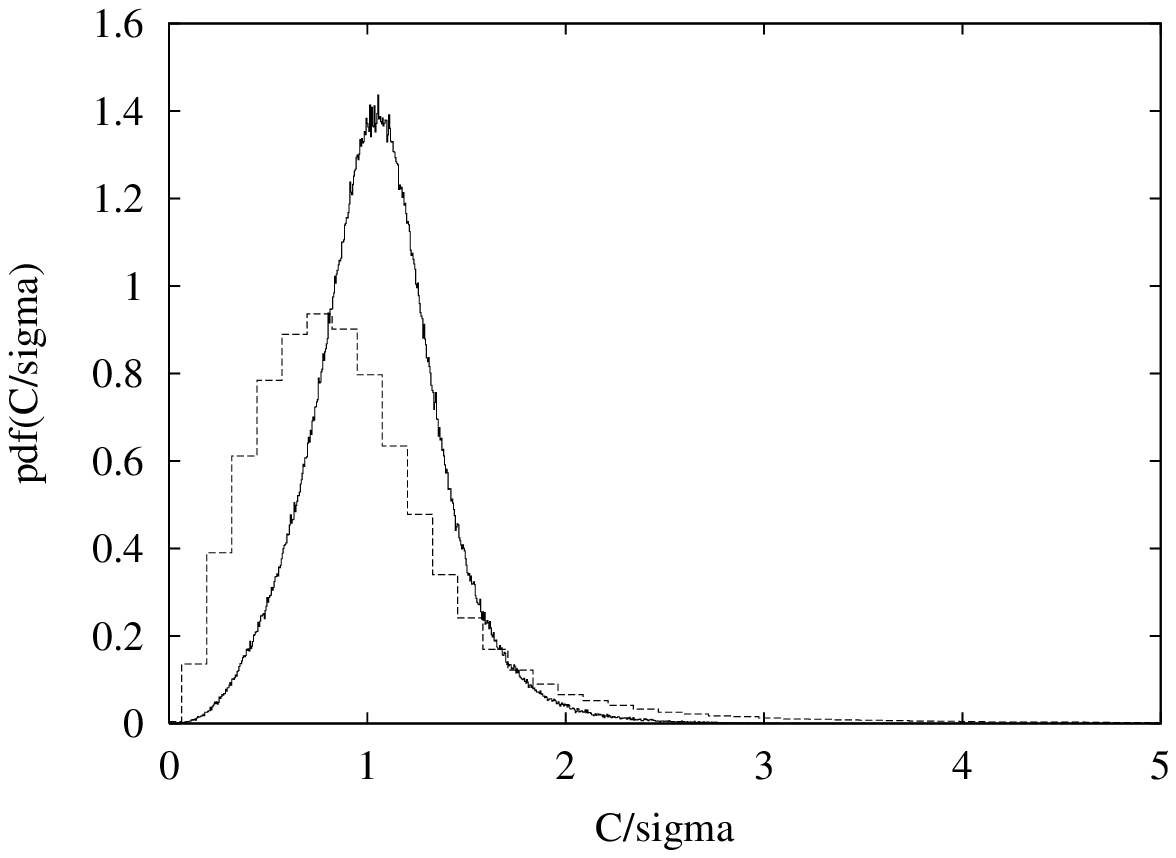}
\includegraphics[width=10cm]{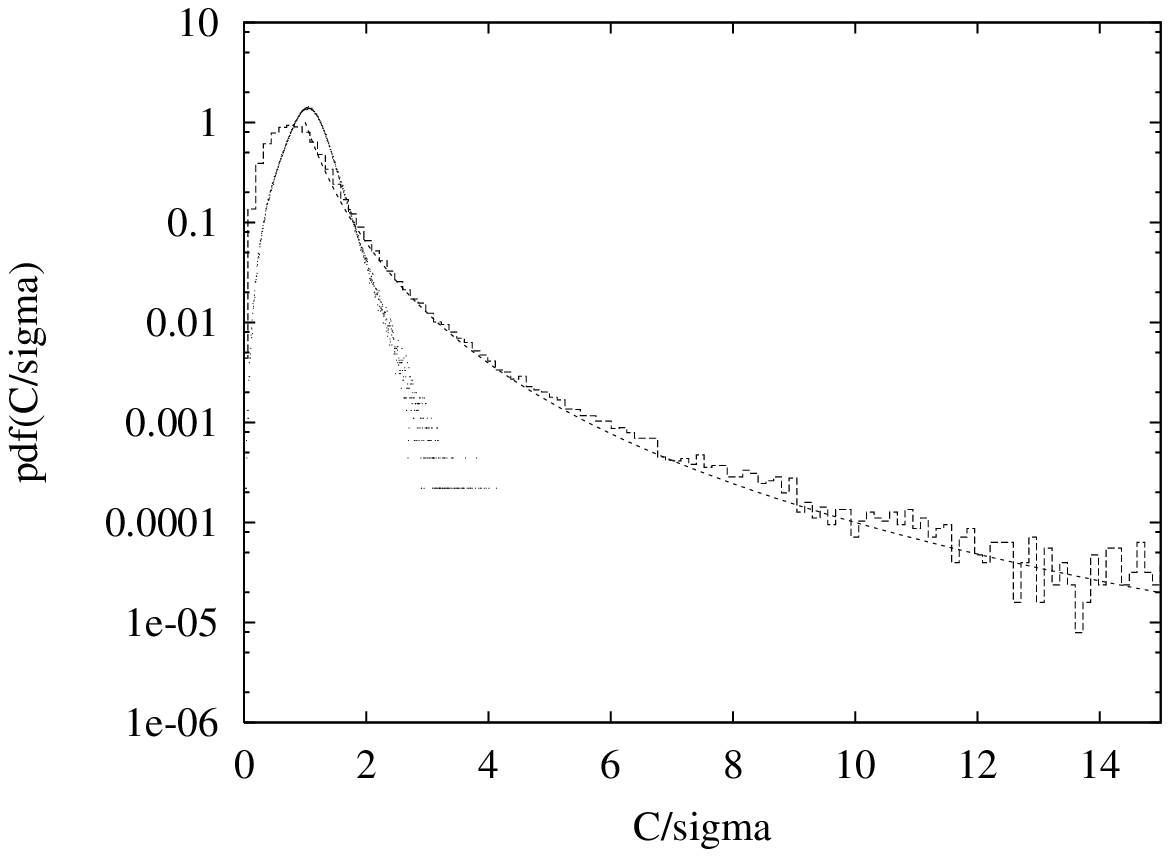}
\parbox[t]{.85\textwidth}
 {
 \caption[fig6]
 {\label{fig6}
\small
The pdf's of $C$, for $N=7$. Top: linear scales, bottom: vertical 
log scale. The distributions with the fatter tails belong to 
Student-t with $m=1$, the remaining stems from Gaussian returns.
 }
 }
\end{center}
\end{figure}

For non-Gaussian distributions, the variance is, of course, not always
an appropriate measure of risk. It is therefore instructive to look at
the full pdf of the quantity $C$. 
Given the hedge functions, it can be easily computed by Monte Carlo. 
In Figure~\ref{fig6} we see
histograms (from 1 million samples of the whole hedging process)
of $C$ ($N=7$) for the Gaussian and the $m=1$ Student-t
returns, both optimally hedged. In the lower part of the figure (with
log scale), we show additionally the function $1/(C/\sigma)^4$,
demonstrating that the power tail of the Student-t distribution is
still around. For comparison: The amplitude in front of the $1/x^4$
term in the 7-fold convolution of the $m=1$ Student-t (with standard
deviation one) is 0.24062.
If one assumes that the option writer would sell the option 
for $\langle C \rangle = 0.96$, his value at risk (VaR(5 \%))
would be 0.98, i.e., more than 100 percent of the option price!


\section{Conclusions}

In this paper, closed form solutions were derived for the optimal 
hedge functions in the case of unbiased Student-t returns. 
I hope that the results prove useful as illustrative 
examples for option pricing in the presence of fat 
tails, and as starting points for further investigations. 

\section*{Acknowledgement} 
Discussions with J. Lemm and C. Wieczerkowski are gratefully
acknowledged. 



\section*{Appendix A: Representations of $\phi_k^*$}

We start from 
$$
\phi_k^*(x) = 
\frac{1}{\sigma^2} \int dr \, p(r) \, r 
\int dt \, P_{N-k-1}(t) \, f(x + r + t) \, , 
$$
with $f(x)= {\rm max}(x-x_s,0)$. The claim function 
can be represented as follows: 
$$ 
f(x)= - \int \frac{dq}{2\pi} \frac{e^{iq(x-x_s)}}{(q-i\epsilon)^2} \, .
$$
To see this one evaluates the integral using the theory of residues.
$$
\int dq \, \frac{e^{iq\lambda}}{(q-i\epsilon)^2} 
= \left. 2 \pi i \, \theta(\lambda) \frac{d}{dq} \, e^{iq\lambda} 
\right\vert_{q=i\epsilon} = - 2 \pi \, \theta(\lambda) \lambda 
= - 2 \pi \, {\rm max}(\lambda,0) \, . 
$$
Here, $\theta(\lambda)$ denotes the Heaviside step function. 
With $x_k - x_s = \xi$ we thus have 
$$
\phi_k^* = 
- \frac{1}{\sigma^2} \int dr \, p(r) \, r \int dt \, P_{N-k-1}(t) 
\int \frac{dq}{2\pi} \, \frac{e^{iq(\xi+r+t)}}{(q-i\epsilon)^2} \, .
$$
Exchanging the order of integration, we obtain 
$$
\phi_k^* = 
- \frac{1}{\sigma^2}
\int \frac{dq}{2\pi} \, \frac{e^{iq\xi}}{(q-i\epsilon)^2}
\int dr \, p(r) \, r \, e^{iqr} 
\int dt \, P_{N-k-1}(t) \, e^{iqt} \, .
$$
The $r$-integral can be written as $\frac1i \frac{d}{dq} \tilde p(q)$, 
whereas the $t$-integral is the same as 
$\tilde p(q)^{N-k-1}$. Putting things together we obtain 
$$ 
\phi_k^* = - \frac{1}{(N-k) \, \sigma^2} 
\int \frac{dq}{2\pi i} \, 
\frac{e^{iq\xi}}{(q-i\epsilon)^2} \, 
\frac{d}{dq} \, \tilde p(q)^{N-k} \, . 
$$
To proceed further, we write $e^{iq\xi}= \cos(q\xi)+i \, \sin(q \, \xi)$. 
Then 
\be 
\label{rrr}
\phi_k^* = - \frac{1}{(N-k) \, \sigma^2} 
\int \frac{dq}{2\pi i} \, 
\frac{\cos{q\xi}}{(q-i\epsilon)^2} \, 
\frac{d}{dq} \, \tilde p(q)^{N-k} \, + \psi^*_k \, ,
\ee 
with 
$$ 
\psi_k^* = - \frac{1}{(N-k) \, \sigma^2} 
\int \frac{dq}{2\pi} \, 
\frac{\sin(q \, \xi)}{q} \, 
\, \frac{1}{q} \, \frac{d}{dq} \, \tilde p(q)^{N-k} \, .
$$ 
In the expression for $\psi^*$ we have already performed 
the limit $\epsilon \rightarrow 0$. 
The first term of eq.~(\ref{rrr}) yields $1/2$. To see this 
we observe that the term is independent of $\xi$: Differentiating
with respect to $\xi$ makes the integrand an odd function of 
$q$ and the integral vanish.
Without changing the integral we can therefore let $\xi=0$.
Furthermore 
$$
\begin{array}{ll}
& - \frac{1}{(N-k) \, \sigma^2} 
\int \frac{dq}{2\pi i} \, 
\frac{1}{(q-i\epsilon)^2} \, 
\frac{d}{dq} \, \tilde p(q)^{N-k} \\[4mm]
&\quad = 
 - \frac{1}{(N-k) \, \sigma^2} \int dx \, P_{N-k}(x) \, x \, 
\int \frac{dq}{2\pi} \, 
\frac{1}{(q-i\epsilon)^2}  \\[4mm]
&\quad = 
 \frac{1}{(N-k) \, \sigma^2} \int_0^\infty  
dx \, P_{N-k}(x) \, x^2 \\[4mm]
&\quad = \frac{1}{2} \, . 
\end{array}
$$
This proves the relation $\phi_k^*= \frac12 + \psi_k^*$. 


\section*{Appendix B: Computation of $\langle f^n \rangle$}

The equation 
$$
\langle f^n \rangle 
= \int  dt \, P_N(t) \, \theta(\zeta + t) \, (\zeta + t)^n 
$$
can be rewritten with the help of the following integral 
representation of the Heaviside step function: 
$$
\theta(x)= 
\int \frac{dq}{2\pi i}
\, \frac{e^{iqx}}{q-i\epsilon} \, , 
$$
where $\epsilon > 0$ is to be considered infinitesimal. 
We obtain 
$$
\langle f^n \rangle 
= 
\int \frac{dq}{2\pi i}
\, \frac{1}{q-i\epsilon} \, 
\int dt \, P_N(t) \, e^{iq(\zeta+t)} \, (\zeta + t)^n \, . 
$$
By elementary operations this can be rewritten as 
$$
\langle f^n \rangle 
= 
\int \frac{dq}{2\pi i}
\, \frac{1}{q-i\epsilon} \, 
\left( \frac1i \frac{d}{dq} \right)^n \, 
\left( e^{iq\zeta} \, \tilde p(q)^N \right) \, .
$$
Working out the derivatives for $n=1,2$, we obtain the representation 
given in eqs.~(\ref{LL1},\ref{LL2}), provided we recognize the  
identities 
\be 
\label{rs1}
\int \frac{dq}{2\pi i}
\, \frac{\cos(q \zeta)}{q-i\epsilon} \, \tilde p(q)^N = \frac12 \, , 
\ee
\be
\label{rs2}
- \int \frac{dq}{2\pi i}
\, \frac{\cos(q \zeta)}{q-i\epsilon} \, 
\frac{d^2}{dq^2} \, \tilde p(q)^N = \frac12 \, N \, \sigma^2 \, , 
\ee
and perform an integration by parts in one of the integrals. 
Eqs.~(\ref{rs1},\ref{rs2}) can 
be proved by first observing that both expressions do 
not depend on $\zeta$. Letting $\zeta=0$ thus leaves 
the integrals invariant.
Transforming back 
to $t$-space we recognize integrals of $P_N(t)$ and $P_N(t) \, t^2$, 
respectively, for $t$ from $0$ to $\infty$.


\section*{Appendix C: Computing $\psi^*$ for Student-t}

We start from eq.~(\ref{psiform}) in the form 
$$
\psi_k^* = \frac{1}{\sigma^2} 
\int_0^{\infty} \frac{dq}{\pi} \, 
\frac{\sin(q \, \xi)}{q} \, 
\, 
\tilde p(q)^{N-k-1} 
\, 
\left( - \frac{1}{q} \, \frac{d}{dq} \, \tilde p(q) \right) \, .
$$ 
It is easy to derive (by partial integration) the following 
recursion relation: 
$$
\frac{1}{q} \, 
\frac{d}{dq} \, 
\tilde p^{(\mu)}(q)  
= 
- \frac{a^2}{\mu - 2} \, 
\tilde p^{(\mu-2)}(q) \, .  
$$
Employing this recurrence and eq.~(\ref{ssig}), we 
obtain 
$$
\psi_k^{[m]*} = 
\int_0^{\infty} \frac{dq}{\pi} \, 
\frac{\sin(q \, \xi)}{q} \, 
e^{-(N-k) \, a  q } \, 
T^{[m]}(aq)^{N-k-1} \,  
T^{[m-1]}(aq) \,  .
$$ 
Performing the substitution $q \rightarrow q/a$ and defining 
$$
\xi' = \frac{\xi}{a} = \frac{x_k-x_s}{a} \, , 
$$
one obtains 
\be
\label{ppp} 
\psi_k^{[m]*} = 
\int_0^{\infty} \frac{dq}{\pi} \, 
\frac{\sin(q \, \xi')}{q} \, 
e^{-(N-k)\, q } \, 
T^{[m]}(q)^{N-k-1} \,  
T^{[m-1]}(q) \,  .
\ee
The product of the two $T$-factors is a polynomial of degree 
$m \, (N-k) - 1$: 
$$
T^{[m]}(q)^{N-k-1} \,  
T^{[m-1]}(q) \equiv \sum_{l=0}^{m(N-k)-1} A^{[m,N-k]}_l \, q^l \, .
$$
E.g., for $m=1$, the $A$-coefficients are given by 
$$
A^{[1,N-k]}_l = 
\left(
\begin{array}{c}
N-k-1 \\
l
\end{array}
\right) \, . 
$$
For larger $m$ the expressions become 
more involved. They can however easily be obtained with 
the help of a computer algebra program. 

Eq.~(\ref{ppp}) becomes 
$$
\psi_k^{[m]*} = \frac1\pi 
\sum_{l=0}^{m(N-k)-1} A^{[m,N-k]}_l 
\int_0^{\infty} dq  \, 
\frac{\sin(q \, \xi')}{q} \, 
e^{-(N-k)\, q } \,  q^l 
$$
This integral can be solved in closed form (Gradstein/Ryshik, 3.944): 
\be
\label{Idef} 
I(\alpha,\beta,\gamma) = \int_0^\infty 
dx \, 
x^{\alpha-1} \, e^{-\beta \,x} \, 
\sin( \gamma x) = \frac{\Gamma(\alpha)}{(\beta^2+\gamma^2)^{\alpha/2}}
\, \sin\left( \alpha \, \arctan \frac\gamma\beta \right) \, .
\ee
We have to identify $\alpha=l$, $\beta= N-k$, and $\gamma=\xi'$. 
Observing that 
$$
\lim_{\alpha \rightarrow 0} I(\alpha,\beta,\gamma) = 
\arctan \frac\gamma\beta  \, , 
$$
we finally arrive at 
$$
\psi_k^{[m]*} = \frac1\pi \arctan \frac{\xi'}{N-k}
+ \sum_{l=1}^{m(N-k)-1} \frac{A_l^{[m,N-k]} \, (l-1)!}{\pi} \, 
\frac{ \sin\left(l \, \arctan\left(\frac{\xi'}{N-k}\right)\right)}
{\left(\xi'^2 + (N-k)^2\right)^{l/2}} \, .
$$
Note that this can alternatively be represented as 
$$ 
\psi_k^{[m]*} = \frac1\pi \arctan \frac{\xi'}{N-k}
+ \sum_{l=1}^{m(N-k)-1} \frac{A_l^{[m,N-k]} \, (l-1)!}{\pi} \, 
\frac{ {\rm Im}\left(N-k + i\, \xi'\right)^l }
{\left(\xi'^2 + (N-k)^2\right)^{l}} \, .
$$ 
One might wish to expand the imaginary part 
$$
{\rm Im}(a+ib)^l = \sum_{j=0}^{[(l-1)/2]}
\left(
\begin{array}{c}
l \\
j
\end{array}
\right) \, (-1)^j \, a^{l-2j-1} \, b^{2j+1} \, , 
$$
thus demonstrating that the corrections to the 
leading $\arctan$ behaviour decay like $\xi'^{-l}$ 
for large $\xi'$. $[j]$ denotes the largest integer 
smaller or equal to $j$. 

\end{document}